%% file: main.tex
\documentclass[10pt,twocolumn,letterpaper]{article}

\usepackage{cvpr}              

\input{preamble}

\definecolor{cvprblue}{rgb}{0.21,0.49,0.74}
\usepackage[pagebackref,breaklinks,colorlinks,allcolors=cvprblue]{hyperref}

\def\confName{CVPR}
\def\confYear{2026}

\title{Residual Primitive Fitting of 3D Shapes with \primitivesname}

\author{
Aditya Ganeshan\footnotemark[1] \quad
Matheus Gadelha\footnotemark[2] \quad
Thibault Groueix\footnotemark[2] \quad
Zhiqin Chen\footnotemark[2] \\
Siddhartha Chaudhuri\footnotemark[2] \quad
Vladimir G. Kim\footnotemark[2] \quad
Wang Yifan\footnotemark[2] \quad
Daniel Ritchie\footnotemark[1]
\\[0.6em]
\footnotemark[1]\, Brown University
\qquad
\footnotemark[2]\, Adobe Research
\\[0.3em]
{\tt\small adityaganeshan@gmail.com}
}

\begin{document}

\twocolumn[{%
\renewcommand\twocolumn[1][]{#1}%
\maketitle
	\centering
	\includegraphics[width=1.0\linewidth]{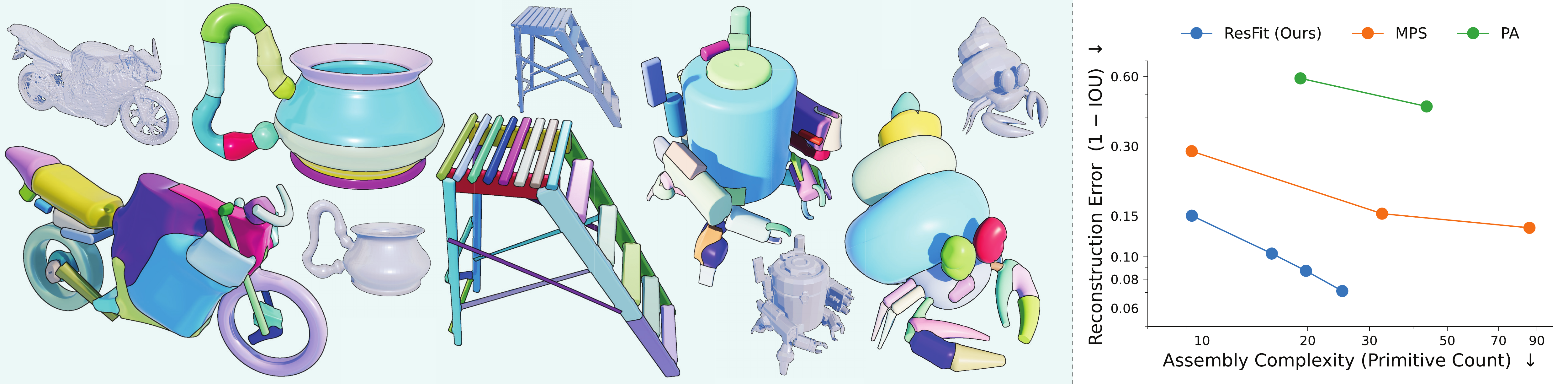}
\captionof{figure}{
(Left) Primitive assemblies inferred by our method capture a wide range of shapes, including hollow forms (vase), curved \& toroidal parts (bike), intricate geometry (ladder, robot), and smooth organic shapes (crab).  
(Right) Our approach shifts the reconstruction–parsimony Pareto frontier: compared to state-of-the-art methods, Marching Primitives~\cite{marchingprim_Liu_2023CVPR} (MPS) and Primitive Anything~\cite{primitiveanything_ye_2025} (PA), we achieve markedly lower reconstruction error using significantly fewer primitives.
}
\vspace{1.5em}
\label{fig:teaser}
}]

\begin{abstract}
We introduce a framework for converting 3D shapes into compact and editable assemblies of analytic primitives, directly addressing the persistent trade-off between reconstruction fidelity and parsimony. 
Our approach combines two key contributions: a novel primitive, termed {\primitivename}, and an iterative fitting algorithm, {\fitname (\fitnameshort)}. 
\primitivename is an analytical primitive that is simultaneously (1) expressive, being able to model various common solids such as cylinders, spheres, cones \& their tapered and bent forms, (2) editable, being compactly parameterized with 8 parameters, and (3) optimizable, with a sign distance field differentiable w.r.t. its parameters almost everywhere. 
\fitnameshort is an unsupervised procedure that interleaves global shape analysis with local optimization, iteratively fitting primitives to the unexplained residual of a shape to discover a parsimonious yet accurate decompositions for each input shape. 
On diverse 3D benchmarks, our method achieves state-of-the-art results, improving IoU by over 9 points while using nearly half as many primitives as prior work. 
The resulting assemblies bridge the gap between dense 3D data and human-controllable design, producing high-fidelity and editable shape programs.
\end{abstract}
\input{sections/011_introduction}

\input{sections/02_related_works}  
\input{sections/031_method}
\input{sections/04_experiments}

\input{sections/05_applications}
\input{sections/06_conclusions}
{
    \small
    \bibliographystyle{ieeenat_fullname}
    \bibliography{main}
}


\end{document}

%% file: preamble.tex
\usepackage{subcaption}

\usepackage{xspace}
\usepackage{listings}
\usepackage{tikz}
\usetikzlibrary{arrows.meta}
\usepackage[ruled,vlined]{algorithm2e}
\usepackage{xcolor}
\definecolor{tealC}{RGB}{0,128,128}
\newcommand{\C}[1]{\textcolor{tealC}{\#~#1}}
\SetKwComment{Comment}{\C{}}{}
\DontPrintSemicolon
\usepackage{amsmath,amssymb}

\definecolor{codeblue}{rgb}{0.25,0.5,0.5}

\lstdefinestyle{algoteal}{
  language=C++,
  backgroundcolor=\color{white},
  basicstyle=\fontsize{7.2pt}{7.2pt}\ttfamily\selectfont,
  columns=fullflexible,
  breaklines=true,
  captionpos=b,
  commentstyle=\fontsize{7.2pt}{7.2pt}\color{codeblue}\ttfamily,
  keywordstyle=\fontsize{7.2pt}{7.2pt}\ttfamily,
  showstringspaces=false,
  mathescape=true
}

\newenvironment{packed_enumerate}
{\begin{enumerate}
    \vspace{-\topsep}
    \setlength{\itemsep}{1pt}
    \setlength{\parskip}{0pt}
    \setlength{\parsep}{0pt}
}{\end{enumerate}}

\definecolor{goldcell}{RGB}{255, 245, 200}
\definecolor{silvercell}{RGB}{235, 235, 235}

\usepackage{pifont}
\usepackage[table]{xcolor}

\usepackage{xspace}
\usepackage{enumitem}
\usepackage{wrapfig}
\usepackage{booktabs,multirow,adjustbox,tabularray}
\usepackage{mathtools,nicefrac}
\usepackage{makecell}
\usepackage{capt-of}
\usepackage[ruled,vlined]{algorithm2e}
\SetAlgoNlRelativeSize{0} 

\usepackage{etoolbox}
\providetoggle{beginFlag}
\settoggle{beginFlag}{true}

\SetAlgoSkip{AlgoSkipBeforeAndAfter}

\usepackage[dvipsnames]{xcolor}
\newcommand{\silenced}[1]{}

\usepackage{dsfont}

\newif\ifshowedits
\newcommand{\addeditor}[3]{%
  \definecolor{#1color}{rgb}{#3}
  \expandafter\newcommand\csname #1\endcsname[1]{%
  \ifshowedits
    {\color{#1color} ##1}%
  \else
    {##1}%
  \fi
  }%
  \expandafter\newcommand\csname #1rmk\endcsname[1]{%
  \ifshowedits
    {\color{#1color} {\bf [#2: ##1]}}
  \fi
  }%
  \expandafter\newcommand\csname #1rpl\endcsname[2]{%
  \ifshowedits
    {{\color{#1color} ##1} \sout{##2}}
  \else
    {##1}
  \fi
  }%
}

\definecolor{darkgreen}{RGB}{0,110,0}
\definecolor{darkred}{RGB}{170,0,0}

\newcommand{\primitivename}{{SuperFrustum}\xspace}
\newcommand{\primitivesname}{{SuperFrusta}\xspace}

\newcommand{\fitname}{{Residual Primitive Fitting}\xspace}
\newcommand{\fitnameshort}{{ResFit}\xspace}

\newcommand{\cmark}{\ding{51}} 
\newcommand{\xmark}{\ding{55}} 

\addeditor{yifan}{YW}{0.7, 0.0, 0.7}

\DeclareMathOperator*{\argmax}{arg\,max}

\showeditstrue

%% file: sections/011_introduction.tex
\begin{figure*}[t!]
	\centering
	\includegraphics[width=1.00\linewidth]{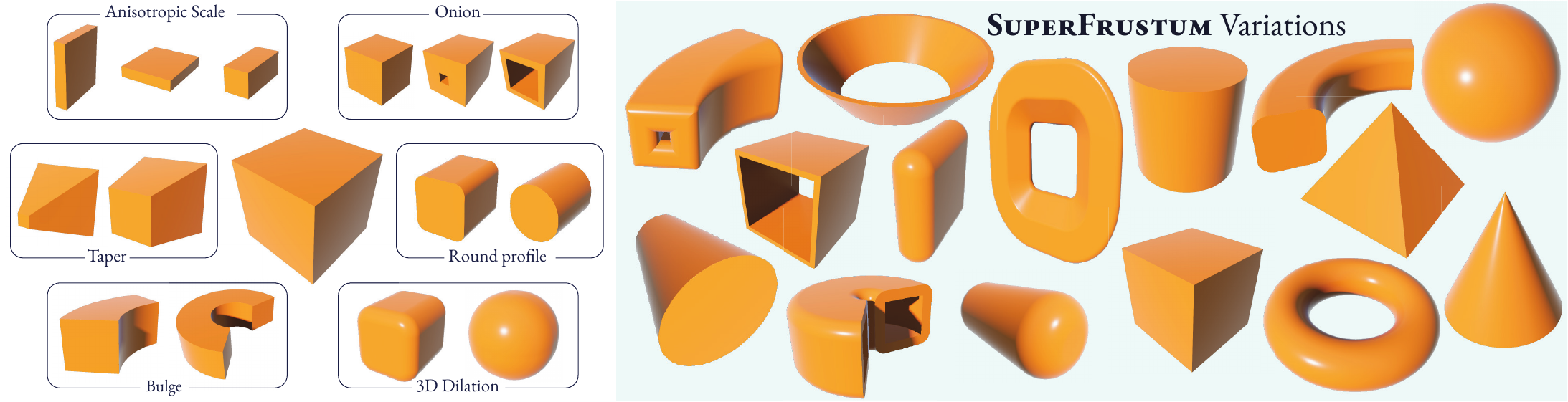}
	\caption{
    \textbf{\primitivename --- An Expressive, Compact \& Differentiable Primitive.}
\primitivename~is a unified analytic SDF primitive with only \textbf{8} parameters controlling dilation, taper, bulge, onion-like hollowing, profile roundness, and axial scaling.  
Its SDF is $C^0$-continuous and fully differentiable (almost eveywhere) with respect to all parameters, enabling robust inverse modeling and gradient-based optimization.  
As shown on the right, these parameters allow a single formulation to morph smoothly across common solids—cuboids, cylinders, cones, spheres, and toroidal variants—and to produce more complex shapes such as bent, hollow, or smoothly capped forms.
}
\label{fig:superprimitive}
\end{figure*}

\section{Introduction}
Recent breakthroughs in 3D generation have enabled the creation of high-quality assets from simple prompts~\cite{liu2024comprehensive,wang2025diffusion}. However, while visually impressive, these outputs are often structurally unorganized, posing challenges for downstream applications like animation, rigging, and interactive editing. Primitive-based representations offer a compelling alternative by distilling complex geometry into a compact assembly of interpretable, analytic parts. This approach yields editable assets and aligns with cognitive findings that humans perceive objects as compositions of simpler forms~\cite{biederman1987recognition}, providing a structured understanding that dense representations lack. The central challenge is converting these unstructured 3D assets into meaningful, primitive-based designs.

Inferring a primitive assembly from a raw 3D shape, however, presents a fundamental trade-off between reconstruction fidelity and program parsimony. Approaches that prioritize high fidelity often yield dense, redundant assemblies of overlapping primitives~\cite{paschalidou2021neural,chen2020bsp,bao20253d,ldif_genova_2020,tertikas2023generating}. Conversely, methods that enforce parsimony may fail to capture fine geometric details or curved structures~\cite{lightsq_wang_2025,marchingprim_Liu_2023CVPR, Paschalidou2019CVPR}. Achieving a representation that is simultaneously expressive, compact, and editable remains an open challenge.

This persistent trade-off can be attributed to two factors. First, commonly used primitive families such as cuboids, superquadrics, or ellipsoids \cite{superquadrics_barr_1981} may require a large number of instances to model the rich shape variations in 3D assets. Second, the inference procedures themselves have distinct limitations. Methods that first commit to a complete segmentation of the input rely on a fixed partition that may not align with what the primitives can efficiently represent~\cite{spheremesh_thiery_2013,generalizedcylinder_zhu_2024,coacd_wei_2022, vhacd_Mammou_2022}. This makes the process brittle, as any initial segmentation errors propagate directly to the final assembly. On the other hand, optimization-driven approaches that fit a large "soup" of primitives from scratch must navigate a highly non-convex loss landscape~\cite{msd_pitas_1990,shapeabstraction_tulsiani_2017,chen2020bsp}.

To address these limitations, we introduce a framework that marries a highly expressive primitive with a robust inference strategy.  
At the core of our approach is the \textbf{\textsc{\primitivename}}, an analytic primitive that fills a key gap in prior work: existing primitive families typically satisfy only one or two of the critical desiderata—\textit{expressivity}, \textit{editability}, and \textit{optimizability}.  
In contrast, \primitivename\ (1) spans common solids such as cylinders, cones, spheres, and their tapered or bent variants; (2) is compactly parameterized with just 8 parameters; and (3) admits a signed-distance field that is differentiable with respect to all parameters, enabling smooth blending and effective inverse modeling.  
Intriguingly, its design builds on analytic functions uncovered by the Shadertoy and Demoscene communities in their pursuit of highly expressive analytic forms with minimal description length~\cite{paniq2017sdUberprim, paniq2017sdSuperprim, paniq2024sdSuperpill, QuilezDistFunctions}.
We find that, when carefully adapted, these formulations are exceptionally well-suited for inverse modeling.

To achieve parsimonious assemblies, an expressive primitive must be paired with an equally effective inference algorithm. We propose \textbf{\textsc{\fitname (\fitnameshort)}}, an unsupervised procedure that tightly interleaves global shape analysis with local primitive optimization to better navigate the highly non-convex reconstruction loss. Instead of optimizing a large set of primitives jointly from scratch, \fitnameshort first analyzes the input geometry to propose initial structures based on global cues. These primitives are then refined via gradient descent to conform to the local geometry. The resulting assembly is subtracted from the target shape, and the process repeats on the unexplained residual. By alternating between proposing global structure and optimizing local parameters, \fitnameshort allows these two signals to mutually inform each other, producing assemblies that are both compact and high-fidelity.

Our approach sets a new state-of-the-art on diverse 3D benchmarks. It consistently produces higher-fidelity reconstructions---\textbf{improving IoU by over 9 points}---while using nearly half the primitives of prior work, demonstrating a fundamental shift in the fidelity-parsimony frontier. These results are enabled by our two primary contributions:

\begin{packed_enumerate}
    \item \textbf{The \primitivename:} A single compact analytic primitive that spans a wide range of canonical volumetric forms while remaining differentiable and suitable for gradient-based optimization.

    \item \textbf{\fitname (\fitnameshort):} An unsupervised inference procedure that alternates between global shape analysis and local primitive optimization to produce compact and accurate assemblies.
\end{packed_enumerate}

Beyond reconstruction, we demonstrate how this framework enables downstream applications including the generation of editable assets, the inference of structured CSG programs, and the enrichment of semantic part segmentations.
\textbf{Code will be open-sourced upon acceptance.}

%% file: sections/02_related_works.tex
\section{Related Works}
\noindent\textbf{Inferring primitive assemblies.}
Existing approaches fall into three main categories.
\emph{Shape-analysis–driven methods}~\cite{spheremesh_thiery_2013,generalizedcylinder_zhu_2024,coacd_wei_2022, vhacd_Mammou_2022, lightsq_wang_2025} partition a shape into regions using geometric cues—such as curvature, thickness, or convexity—and then fit primitives to these regions.  
They produce structurally coherent decompositions when the partitions match the primitive family, but are often brittle across diverse shapes and sensitive to tuning.  
Since the decomposition is fixed and independent of what the primitives can represent, these methods struggle to balance fidelity and compactness.
\emph{Optimization-driven methods} directly adjust primitive parameters to minimize reconstruction error for a target shape~\cite{msd_pitas_1990, marchingprim_Liu_2023CVPR} or its renders~\cite{monnier2023dbw}.  
While effective on small assemblies, they often require many primitives for high fidelity, as reconstruction loss tends to dominate disentanglement and parsimony without strong initialization.
\emph{Learned methods} predict primitive parameters or part layouts using neural networks. Some methods train the network on supervised data~\cite{primitiveanything_ye_2025, grass, shape_assembly, mo2019structurenet} while others formulate unsupervised reconstruction-based objectives to infer the assemblies~\cite{ldif_genova_2020,chen2020bsp, 8237365, vpie_jones_2024, progrip_deng_2022,unsup_cube_hier, split_merge_refine, unsup_cuboid, deng2020cvxnet, paschalidou2021neural,Paschalidou2019CVPR}. 
Such models achieve high reconstruction accuracy on domains similar to their training data but generalize poorly to novel or complex objects.

Our approach combines the strengths of analysis- and optimization-driven paradigms: rather than committing to a single decomposition or a fixed primitive set, we \emph{interleave} analysis and optimization so that each informs the other.  
This bidirectional formulation adapts the decomposition to the representational capacity of the primitives, producing assemblies that remain both compact and geometrically faithful.

\noindent\textbf{Primitive representations.}
Primitive design has progressed from simple analytic forms to more expressive but increasingly complex parameterizations.  
Early methods used cuboids or cylinders~\cite{shapeabstraction_tulsiani_2017,ldif_genova_2020}, which are interpretable but limited in expressivity.  
Superquadrics~\cite{superquadrics_barr_1981, superdec_fedele_2025, Paschalidou2019CVPR} and algebraic surfaces~\cite{Yavartanoo_2021_ICCV} enlarge the shape space but sacrifice editability and cannot exactly reproduce canonical solids such as cubes or cones—common in manufactured objects.  
Recent generalized-cylinder–based primitives~\cite{zhao2024sweepnet, extrusionprimitive_Chunyi_2025} increase flexibility yet still fall short in reconstruction fidelity.  
Neural implicit part representations~\cite{ldif_genova_2020, paschalidou2021neural, 10.5555/3600270.3601383,speg} offer high expressivity but are opaque, costly, and difficult to control or reuse.

In parallel, the graphics and demoscene communities have explored unified analytic primitives that morph between basic shapes within a single functional form~\cite{QuilezDistFunctions,paniq2024sdSuperpill,paniq2017sdSuperprim,paniq2017sdUberprim}.  
These formulations were developed to minimize scene-description size and enable real-time rendering, not for inverse modeling or differentiable fitting.  
\primitivename\ draws inspiration from this work, extending it to a broader shape space and demonstrating its effectiveness for high-fidelity primitive assembly inference.

%% file: sections/031_method.tex
\section{Method}

\begin{figure}[t!]
	\centering
	\includegraphics[width=1.00\linewidth]{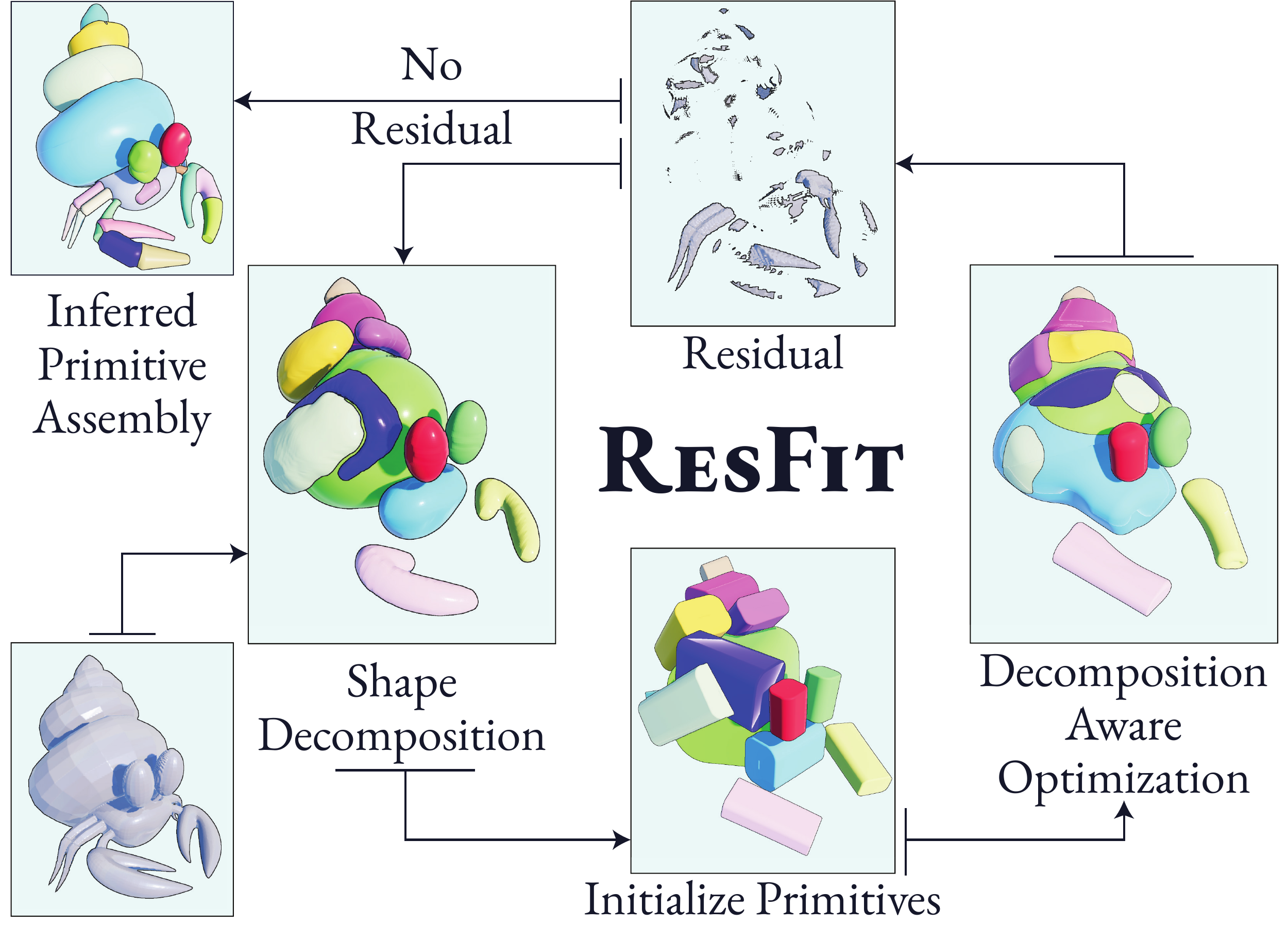}
	\caption{ 
    \textbf{\fitnameshort}~infers parsimonious assemblies by interleaving \textit{shape analysis} and \textit{primitive optimization}.  
Shape decomposition provides initial primitives, which are refined with decomposition-aware optimization.  
Residual unexplained volumes are then extracted and seeded with new primitives.  
 }
\label{fig:resfit}
\end{figure}

We define the primitive assembly inference task as follows:  
given a 3D shape $x$, our goal is to infer a primitive assembly $z$ composed of analytic primitives whose execution $E(z)$ reconstructs the input shape.  
Each program $z$ defines a sequence of primitives $\{f_{\theta_i}\}_{i=1}^{|z|}$ combined through compositional operators to yield a closed surface $E(z)$.  
Following Occam’s razor, we seek programs that are both \textit{accurate} and \textit{compact}.  
Formally, we aim to maximize the following objective:
\begin{align}
\label{eq:objective_main}
    z^* &= \argmax_{z}\; \mathcal{O}(x, z), \\
\label{eq:objective_terms}
    \mathcal{O}(x, z) &= \mathcal{R}(x, E(z)) - \alpha |z|,
\end{align}
where $\mathcal{R}$ measures the reconstruction accuracy between the input shape $x$ and the program execution $E(z)$, 
$|z|$ denotes the program complexity (or number of primitives in the program), and $\alpha$ controls the trade-off between accuracy and compactness.  
Maximizing $\mathcal{O}$ thus favors concise programs that explain the geometry with a small set of expressive parts.  

We now summarize the components of our method.
Section~\ref{sec:fiting} introduces \fitnameshort, our iterative fitting procedure.
Section~\ref{sec:superprimitive} defines \primitivename, the unified analytic primitive used in all assemblies.
Section~\ref{sec:msd-init} describes our MSD-based initialization strategy, and Section~\ref{sec:prog-opt} details the optimization process that balances geometric fidelity with parsimony.

\subsection{\textbf{\fitnameshort~: \fitname}}
\label{sec:fiting}

\begin{figure*}[t!]
	\centering
	\includegraphics[width=1.00\linewidth]{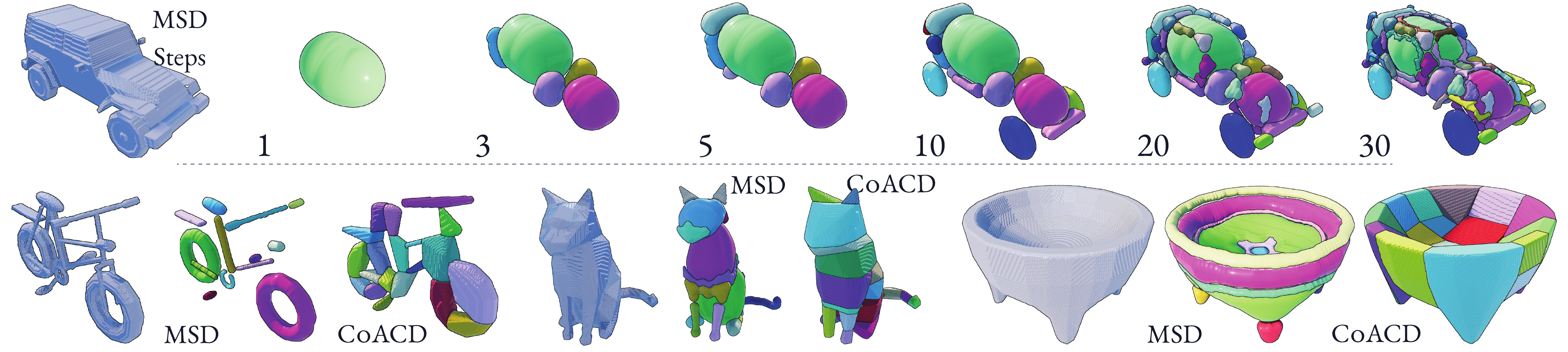}
	\caption{ 
    Morphological Shape Decomposition (MSD) iteratively extracts connected regions of similar thickness.  
Top: successive MSD partitions of a input mesh.  
Bottom: MSD yields regions that form suitable initialization seeds for \primitivesname—capturing non-convex structures such as bicycle tires (left), a cat’s curved tail (center), and bowl rims (right).  
In contrast, CoACD over-partitions these regions into many convex fragments, often using axis-aligned cuts that produce semantically misaligned parts.}

\label{fig:MSD}
\end{figure*}

Purely optimization-based methods often produce entangled reconstructions, while analysis-based approaches partition shapes without considering the primitive family's representational capacity. 
This creates a disconnect between the geometric analysis (top-down) and the primitive representation (bottom-up). 
\fitname~(\fitnameshort) bridges this divide by interleaving shape analysis and assembly optimization, allowing each phase to inform the other and yielding assemblies that are both compact and geometrically faithful.

Our procedure alternates between analysis and optimization (Fig.~\ref{fig:resfit}). 
The analysis stage decomposes the current residual volume into regions that seed new primitives. 
The optimization stage then adjusts parameters to maximize $\mathcal{O}$ (Eq.~\ref{eq:objective_main}), separating explained geometry from remaining residuals. 
This cycle repeats until $\mathcal{O}$ saturates or a fixed iteration budget $K$ is reached.

Several design choices ensure that the iterative loop can correct both over- and under-parameterization.  
To prevent over-parameterization, we seed few primitives per round and employ parsimony-aware optimization: a soft regularizer penalizes redundancy during fitting, while hard pruning removes parts that degrade $\mathcal{O}$. 
To address under-parameterization, primitives are optimized based on their local support, and the full assembly is re-optimized in each round. 
This enables self-correction as new parts are added, allowing the system to converge toward a compact and coherent structure.

\subsection{Expressive, Editable \& Optimizable Primitive}
\label{sec:superprimitive}
\setlength{\columnsep}{3pt}  
\setlength{\intextsep}{4pt}  
\begin{wrapfigure}{r}{0.67\linewidth} 
    \vspace{-0.5\baselineskip}       
    \centering
    \footnotesize
    \setlength{\tabcolsep}{1pt}      
    \begin{tabular}{@{}lccc@{}}      
        \toprule
        Primitive & Expressive & Editable & Optimizable \\
        \midrule
        S. quadrics~\cite{Paschalidou2019CVPR}
            & \xmark & \cmark & \cmark \\
        Alg. Surf~\cite{Yavartanoo_2021_ICCV}
            & \cmark & \xmark & \cmark \\
        Multi-type~\cite{primitiveanything_ye_2025}
            & \cmark & \cmark & \xmark \\
        \specialrule{0.5pt}{0.25pt}{0.25pt}
        \rowcolor{goldcell}
        \textbf{\primitivesname} & \cmark & \cmark & \cmark \\
        \bottomrule
    \end{tabular}    
    \label{tab:primitive-criteria}
    \vspace{-0.5\baselineskip}       
\end{wrapfigure}
An ideal primitive for inverse graphics must be \emph{expressive} enough for diverse forms, \emph{editable} via intuitive controls, and robustly \emph{optimizable}. As existing families often fall short, we introduce \primitivesname, a unified analytic primitive designed to meet all three desiderata.

A \primitivename\ is the zero-level set of a signed distance function 
\begin{equation}
SF(\mathbf{p}) \;=\; f(\mathbf{p}; \theta),
\qquad
\theta \;=\; (\mathbf{s}, r, d, t, b, o),
\end{equation}
with parameters $\theta = (\mathbf{s}, r, d, t, b, o)$. These 8 scalars intuitively control anisotropic scale ($\mathbf{s}$), profile rounding ($r$), dilation ($d$), taper ($t$), bulge ($b$), and onion/shell thickness ($o$), as shown in Fig.~\ref{fig:superprimitive} (further implementation details and the reference code are provided in the supplementary). Its continuous, piecewise-\(C^1\) formulation spans a wide range of shapes including cuboids, cylinders, cones, and tori, and is differentiable almost everywhere, enabling stable gradient-based fitting. 

The complete primitive assembly $z = E(z)$ is formed by composing transformed \primitivesname. Each instance $i$ has a pose $(R_i, t_i)$ and shape parameters $\theta_i$, yielding a signed distance $g_i(\mathbf{p}) = f(R_i^\top(\mathbf{p} - t_i); \theta_i)$. The final implicit field $\mathcal{F}$ is obtained by recursively applying a smooth union operator $U$:
\begin{equation}
\begin{aligned}
\mathcal{F}_{1}(\mathbf{p}) &= g_{1}(\mathbf{p}), \\
\mathcal{F}_{k+1}(\mathbf{p}) &= U\!\big(\mathcal{F}_k(\mathbf{p}),\, g_{k+1}(\mathbf{p});\, \beta_k\big),
\end{aligned}
\end{equation}
where $\beta_k$ controls blend sharpness. The final surface is the zero level set of $\mathcal{F}$.

\subsection{Shape Decomposition for \primitivesname}
\label{sec:msd-init}

\fitnameshort initializes primitives from the volumetric regions produced by a shape decomposition method, and its performance improves when the chosen decomposition strategy aligns with the primitive family's expressiveness. While recent work adapt Approximate Convex Decomposition (ACD)~\cite{lightsq_wang_2025} for initializing primitives, we find an adapted variant of Morphological Shape Decomposition (MSD)~\cite{msd_pitas_1990} is more suitable for initializing \primitivesname.
 
MSD is an iterative “peel the thickest part first” technique. At each step, it finds the largest connected region of roughly uniform thickness, extracts it, removes it from the shape, and repeats on the residual. This process yields a thickness-ordered set of volumetric regions for primitive initialization, as shown in Figure~\ref{fig:MSD}.

Formally, given a signed distance field $f(\mathbf{p})$, each iteration $k$ identifies the thickest interior region $\Gamma_k$ by finding the connected component (cc) that survives erosion up to a radius $|\tau|$:
\begin{equation}
\Gamma_k \subseteq \{\mathbf{p}\!\in\!\Omega \mid f(\mathbf{p}) \le \tau\}, 
\qquad \Gamma_k \text{ is a cc.}
\end{equation}
The threshold $\tau \le 0$ is the minimum value such that $\mathrm{Vol}(\Gamma_k)$ meets a volume fraction $\kappa$. To recover its full spatial extent, we dilate $\Gamma_k$ back by the same radius, $R_k = \Gamma_k \oplus B_{|\tau|}$. This part $R_k$ is recorded and subtracted from the shape by updating the residual field:
\begin{equation}
f_{k+1}(\mathbf{p}) = f_k(\mathbf{p}) \setminus R_k,
\quad f_1 \equiv f.
\end{equation}
Repeating this process produces a sequence of candidate regions $\{R_k\}$ ordered by decreasing thickness.

MSD offers two key advantages over ACD~\cite{acd_lien_2007, coacd_wei_2022} for this task. First, ACD's convexity constraint over-partitions non-convex structures that a single \primitivename\ can model, such as the bent and hollow forms shown in Figure~\ref{fig:MSD} (bottom). Second, MSD is substantially more robust to the noisy surface artifacts present in the residual volumes generated during our iterative fitting loop, making it better suited for \fitnameshort.

For each decomposed part volume, we instantiate a \primitivename. We initialize its parameters by using PCA on points sampled within the volume. 
Cylindricity score along the different PCA axis is used to select a canonical direction. Pose $(R,t)$ and Scale is then inferred w.r.t the canonical axis. Refer to the supplementary for further details.

\subsection{Decomposition-Aware Optimization}
\label{sec:prog-opt}

We optimize the assembly parameters to maximize the objective $\mathcal{O}$ (Eq.~\ref{eq:objective_main}) in two stages. First, a differentiable phase minimizes a corresponding loss via gradient descent. Second, a discrete pruning phase removes primitives that degrade $\mathcal{O}$. The differentiable loss comprises three components addressing reconstruction fidelity, program parsimony, and program quality.

\noindent
\textbf{Reconstruction.}
The reconstruction loss is a differentiable surrogate for $\mathcal{R}$ in Eq.~\ref{eq:objective_terms}. We supervise the predicted occupancy field $\hat{o}(\mathbf{p}) = \sigma(-\beta\,\tanh\!\big(\beta\,\mathcal{F}(\mathbf{p})))$ of the current assembly against the ground-truth occupancy $o(\mathbf{p})$. Samples $\mathbf{p}$ are drawn uniformly from the shape's volume and densely near its surface. To better reconstruct thin, high-curvature structures, each point is weighted by the principal curvature $\kappa(\mathbf{p})$ of the target mesh. The loss is evaluated only within a spatial mask $\mathcal{M}=\{\mathbf{p}\mid \mathcal{F}(\mathbf{p})<\tau\}$ to focus optimization on signals from the assembly's vicinity.
\begin{equation}
\begin{aligned}
w(\mathbf{p}) &= 1 + \sigma(\kappa(\mathbf{p})), \\
\mathcal{L}_{\text{rec}} &=
\frac{1}{|\mathcal{M}|}
\sum_{\mathbf{p}\in\mathcal{M}}
w(\mathbf{p})
\big(\hat{o}(\mathbf{p}) - o(\mathbf{p})\big)^2.
\end{aligned}
\end{equation}
\noindent
\textbf{Parsimony.}
To encourage compact assemblies, each primitive $i$ is assigned a stochastic existence variable $q_i \in (0,1)$ sampled via a Gumbel-Softmax distribution. Its signed distance field is then modulated as
$
f_i^*(\mathbf{p}) = q_i\,f_i(\mathbf{p}) + (1-q_i),
$
which smoothly erodes primitives with low existence probability. The parsimony loss penalizes the expected number of active primitives:
$
\mathcal{L}_{\text{count}} = \sum_i q_i.
$

\noindent
\textbf{Quality.}
To improve editability and prevent geometrically entangled or overly blended assemblies, we add a structural regularizer that combines overlap and smooth-union consistency losses:
\begin{equation*}
\begin{aligned}
\mathcal{L}_{\text{qual}} 
= 
\underbrace{
\max(1, \sum\nolimits_i  \hat{o}_i(\mathbf{p}))}_{\mathcal{L}_{\text{overlap}}}
+
\underbrace{
\hat{o}(\mathbf{p}) - \min(\sum\nolimits_i  \hat{o}_i(\mathbf{p}), 1)}_{\mathcal{L}_{\text{union}}}
,
\end{aligned}
\end{equation*}
where $\hat{o}_i$ is the occupancy of primitive $i$. The $\mathcal{L}_{\text{overlap}}$ term penalizes regions where multiple primitives are simultaneously active, discouraging redundant coverage. The $\mathcal{L}_{\text{union}}$ term penalizes regions that are occupied by the smooth union assembly but not by any of the independent primitives, discouraging excessive blending.

The total differentiable loss is the weighted sum of these components:
\begin{equation}
\label{eq:overall_loss}
\mathcal{L}_{\text{total}} =
\mathcal{L}_{\text{rec}} +
\lambda_{\text{count}}\mathcal{L}_{\text{count}} +
\lambda_{\text{qual}}\mathcal{L}_{\text{qual}}.
\end{equation}

\noindent
\textbf{Pruning.}
After the differentiable optimization converges, a discrete pruning step further simplifies the assembly. Primitives with negligible volume or contribution are tested for removal, and deletions are greedily accepted if they improve the primary objective $\mathcal{O}$.

%% file: sections/04_experiments.tex
\section{Experiments}
\begin{table*}[t]
\small
\centering
\setlength{\tabcolsep}{10pt}
\begin{tabular}{lcccccccc}
\toprule
\multirow{2}{*}{\textbf{Method}}
& \multicolumn{4}{c}{\textbf{Reconstruction Quality}} &
  \multicolumn{4}{c}{\textbf{Program Quality}} \\
\cmidrule(lr){2-5} \cmidrule(lr){6-9}
 &
{\footnotesize IOU ($\uparrow$)} &
{\footnotesize BiSurfIOU ($\uparrow$)} &
{\footnotesize CD ($\downarrow$)} &
{\footnotesize EMD ($\downarrow$)} &
{\footnotesize \#Prims ($\downarrow$)} &
{\footnotesize Overlap ($\downarrow$)} &
{\footnotesize IntraPrim ($\downarrow$)} &
{\footnotesize InterPrim ($\uparrow$)} \\
\midrule
\multicolumn{9}{c}{\textbf{3DGen-Prim Dataset}} \\
\midrule
PA~\cite{primitiveanything_ye_2025}     & 34.62 & 30.56 & 7.588 & 0.119 & \cellcolor{goldcell}19.81 & 0.402 & 0.359 & \cellcolor{goldcell} 0.238 \\
PA (TTO) & 53.73 & 37.90 & 0.742 & 0.131 & 64.87 & 0.493 & \cellcolor{silvercell} 0.244 & 0.156 \\
MPS~\cite{marchingprim_Liu_2023CVPR} & \cellcolor{silvercell} 82.67 & \cellcolor{silvercell} 71.53 & \cellcolor{silvercell} 0.884 & \cellcolor{silvercell} 0.093 & 42.96 & 0.684 & 0.250 &  0.157 \\
\specialrule{0.5pt}{0.25pt}{0.25pt}
Ours & \cellcolor{goldcell} 88.74 & \cellcolor{goldcell} 80.19 & \cellcolor{goldcell} 0.168 & \cellcolor{goldcell} 0.073 & \cellcolor{silvercell} 23.98 & \cellcolor{goldcell} 0.210 & \cellcolor{silvercell} 0.244 & \cellcolor{silvercell} 0.207 \\
\midrule
\multicolumn{9}{c}{\textbf{Toy4K Dataset}} \\
\midrule
PA~\cite{primitiveanything_ye_2025} & 36.76 & 31.63 & 5.771 & 0.142 & \cellcolor{goldcell}19.14 & 0.341 & 0.322 & \cellcolor{goldcell} 0.255 \\
PA (TTO) & 49.08 & 37.78 & 1.220 & 0.146 & 44.78 & 0.408 & \cellcolor{silvercell} 0.237 & 0.179 \\
MPS~\cite{marchingprim_Liu_2023CVPR} & \cellcolor{silvercell} 80.60 & \cellcolor{silvercell} 72.75 & \cellcolor{silvercell} 1.147 & \cellcolor{silvercell} 0.086 & 30.62 & 0.588 & 0.245 &  0.201 \\
\specialrule{0.5pt}{0.25pt}{0.25pt}
Ours & \cellcolor{goldcell} 89.92 & \cellcolor{goldcell} 85.66 & \cellcolor{goldcell} 0.154 & \cellcolor{goldcell} 0.067 & \cellcolor{silvercell} 23.67 & \cellcolor{goldcell} 0.208 & \cellcolor{goldcell} 0.221 & \cellcolor{silvercell} 0.202 \\
\bottomrule
\end{tabular}
\caption{Evaluation on 3DGen-Prim~\cite{3dgenbench_zhang_2025} and Toys4K~\cite{toy4k_Stojanov_2021} datasets.
Our method achieves the best reconstruction and program quality scores simultaneously—improving IOU by 6–9 points while using roughly half as many primitives.
Gold = best, Silver = second best.}
\label{tab:two_dataset_eval}
\end{table*}

\noindent
\textbf{Datasets.}  
We evaluate on two datasets capturing generated and real-world assets. As the \emph{3DGen-Prim} dataset~\cite{lightsq_wang_2025} is not public, we recreate it using 510 prompts from 3DGen-Bench~\cite{3dgenbench_zhang_2025} with the Hunyuan3D-2.1~\cite{hunyuan3d_tencent_2025} generator. Our second dataset contains 500 geometrically diverse shapes from Toys4K~\cite{toy4k_Stojanov_2021}, selected via farthest-point sampling.

\noindent
\textbf{Metrics.}  
We evaluate \emph{reconstruction accuracy} and \emph{program quality}.
For accuracy, we use standard metrics: voxel IoU ($128^3$), Chamfer Distance (CD), and Earth Mover's Distance (EMD) over 2048 randomly sampled surface points.
We also report a \emph{Bidirectional Surface IoU (BiSurfIoU)} to better capture surface fidelity, computed as the mean of IoU scores from near-surface points sampled on both the target and reconstructed shapes, with the latter surface extracted via dual contouring of the SDF. 
For program quality, we introduce four metrics. Program length ($|z|$) and \emph{Overlap Ratio} (the volumetric percentage of the shape covered by multiple primitives) measure parsimony and redundancy. To quantify semantic coherence, we use two metrics based on PartField features~\cite{partfield_Liu_2025}. After associating surface points on the target mesh to their nearest primitive, we compute: \emph{IntraPrim}, the mean feature variance within each primitive (lower is better), and \emph{InterPrim}, the average nearest-neighbor distance between primitive feature centroids (higher is better). These scores are aggregated using a size-weighted average.

\noindent
\textbf{Baselines.}  
We compare our method against two state-of-the-art approaches. \emph{Primitive Anything (PA)}\cite{primitiveanything_ye_2025} is a learning-based method trained on a large dataset of manually annotated shapes to predict assemblies of cuboids, cylinders, and ellipsoids from point cloud inputs. Following the original work, we also report its test-time optimization variant, \emph{PA (TTO)}, which refines its predictions using Chamfer Distance. \emph{Marching Primitives (MPS)}\cite{marchingprim_Liu_2023CVPR} serves as a strong optimization-based baseline that directly optimizes a superquadric-based assembly from an SDF grid to achieve state-of-the-art reconstruction fidelity. We run MPS at 128 voxel resolution to match our input. 
We omit comparisons to methods outperformed by MPS~\cite{ems_liu_2022, superdec_fedele_2025} or those without public code~\cite{lightsq_wang_2025, extrusionprimitive_wang_2025}.

\noindent
\textbf{Implementation Details.}  
All experiments use a fixed set of hyperparameters unless stated otherwise. \fitnameshort runs for a maximum of 10 fitting rounds or until convergence, with each round applying 7 iterations of MSD. The high-level objective $\mathcal{O}$ (Eq.~\ref{eq:objective_main}) combines curvature-weighted surface IoU with a program-length penalty ($\alpha = 10^{-3}$). During optimization, the loss weights are set to $\lambda_{\text{count}} = 10^{-3}$ and $\lambda_{\text{qual}} = 10^{-2}$ (Eq.~\ref{eq:overall_loss}). Additional optimization details and ablations are provided in the supplementary material.

\subsection{Parsimonious High fidelity Assemblies}

Table~\ref{tab:two_dataset_eval} summarizes the reconstruction and program quality metrics on both datasets. 
Across all reconstruction measures, our method improves IoU scores by $+6.1$ points on 3DGen-Prim and $+9.3$ points on Toys4K over prior work. 
We attribute this performance to the expressivity of the \primitivename primitive and the iterative analysis-optimization loop of \fitnameshort.

These reconstruction gains are accompanied by improved program quality. 
Our assemblies use approximately half as many primitives as Marching Primitives while reducing volumetric overlap by over $3\times$. 
The inferred primitives also demonstrate high semantic coherence: our method achieves the lowest \emph{IntraPrim} scores, indicating high semantic purity within primitives, and among the highest \emph{InterPrim} scores, reflecting meaningful distinctions between parts. 
These results demonstrate that \fitnameshort produces assemblies that are simultaneously more \textit{accurate}, \textit{compact}, and \textit{semantically interpretable}.

\begin{figure*}[t!]
	\centering
	\includegraphics[width=1.00\linewidth]{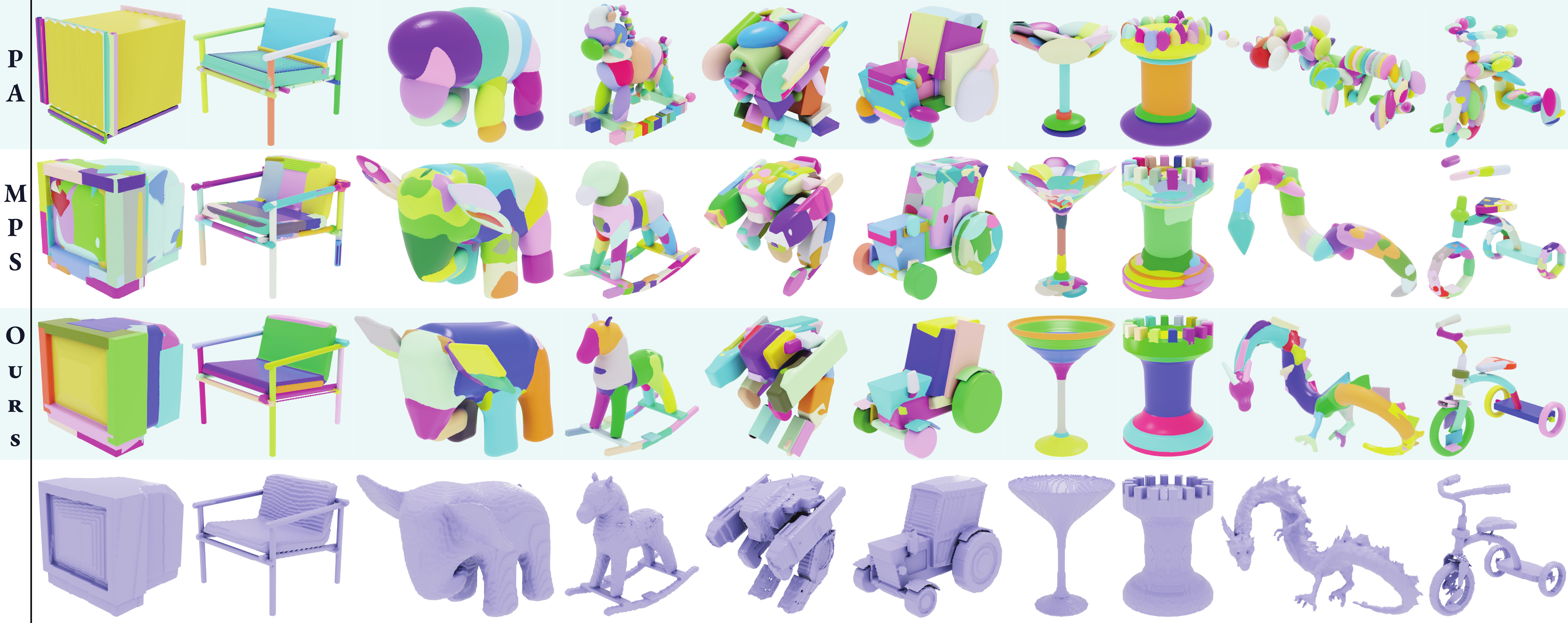}
	\caption{
    Our method reconstructs target shapes with high geometric fidelity and produces more interpretable assemblies, using compact, minimally-overlapping primitives. 
In contrast, Primitive Anything~\cite{primitiveanything_ye_2025} (PA) and Marching Primitives~\cite{marchingprim_Liu_2023CVPR} (MPS) often lose fine structure and generate assemblies with substantial primitive overlap.
 }
\label{fig:qualitative}
\end{figure*}

Qualitative comparisons in Figure~\ref{fig:qualitative} corroborate these findings. 
Our assemblies exhibit higher geometric fidelity and are more interpretable, using a compact set of non-overlapping, semantically aligned primitives. 
In contrast, baseline reconstructions can show lower fidelity on complex structures and tend to produce assemblies with greater primitive overlap.

\subsection{Ablative Analysis}
\label{subsec:ablation}
\noindent
\textbf{Primitive Assembly Design.}
Table~\ref{tab:repr-ablation} compares different primitive families and composition operators. Our full \primitivename~formulation achieves the highest reconstruction fidelity. Disabling smooth unions reduces accuracy and increases overlap, as continuous volumes must then be formed by intersecting primitives rather than by smooth blending. Superprimitive~\cite{paniq2017sdSuperprim}, a variant of our primitive without tapering or bending also lowers accuracy, confirming these degrees of freedom are important for capturing curved and non-uniform structures. Substituting our primitive with cuboids or superquadrics (SQs) further degrades performance. SQs are particularly susceptible to poor local minima when their axes misalign with the target geometry—an issue that methods like MPS mitigate via non-differentiable heuristics such as periodic axis-flipping, which are excluded from our controlled comparison.

\begin{table}[t]
\small
\centering
\setlength{\tabcolsep}{6pt}
\renewcommand{\arraystretch}{1.15}
\begin{tabular}{cc
ccccc}
\toprule
 & \textbf{S. Union} & {\footnotesize IOU}  & {\footnotesize CD}  & {\footnotesize \#Prims}  & {\footnotesize Overlap} \\
\midrule
{\small Cuboid} & \xmark & 82.33 & \cellcolor{goldcell}0.129 & 20.49 & 0.298 \\
{\small S Q}     & \xmark & 76.50 & 0.523 & \cellcolor{goldcell}18.17 & 0.286 \\
{\small S P}     & \xmark & \cellcolor{silvercell}86.66 & 0.134 & 21.36 & 0.291 \\
\specialrule{0.5pt}{0.25pt}{0.25pt}
\textsc{S F}            & \xmark & 87.15 & 0.173 & \cellcolor{silvercell}18.68 & \cellcolor{silvercell}0.257 \\
\textsc{S F}          & \cmark & \cellcolor{goldcell}88.37 & \cellcolor{silvercell}0.147 & 21.46 & \cellcolor{goldcell}0.199 \\
\bottomrule
\end{tabular}
\caption{
Primitive representation ablation: 
\primitivename delivers superior performance over Cuboids, Superquadrics (SQ) and SuperPrimitive(SP) (ref. Section~\ref{subsec:ablation}). Combining \primitivename with smooth union further improves performance.
}
\label{tab:repr-ablation}
\end{table}

\noindent
\textbf{Decomposition and Fitting Strategy.}  
Table~\ref{tab:decompose-ablation} compares our iterative \fitnameshort procedure against a single-shot fitting baseline, using both MSD and CoACD for initialization. The single-shot approach optimizes all primitives simultaneously after the initial decomposition. This makes it sensitive to the initial partition, as it has no mechanism to reallocate capacity to unexplained regions, resulting in less accurate and less compact assemblies.

In contrast, \fitnameshort uses multiple refinement rounds to progressively reallocate primitives toward residual errors and prune where unnecessary. This iterative process achieves higher fidelity with fewer primitives and lower overlap. When comparing decomposition strategies, MSD consistently outperforms CoACD. MSD's ability to produce non-convex partitions provides better initializations for our primitives, especially on the curved, hollow, and branching geometries as shown in Figure~\ref{fig:MSD}.

\begin{table}[t]
\small
\centering
\setlength{\tabcolsep}{6pt}
\renewcommand{\arraystretch}{1.15}
\begin{tabular}{cccccc}
\toprule
\textbf{Step} & \textbf{Method} &
{\footnotesize IOU}  &
{\footnotesize CD}  &
{\footnotesize \#Prims} &
{\footnotesize Overlap} \\
\midrule
{\small Single} & {\small CoACD} & 86.56 & 0.368 & 26.67 & 0.226\\
{\small Single} & {\small MSD}   & 87.95 & 0.337 & 28.17 & 0.236 \\
{\small \fitnameshort}  & {\small CoACD} & \cellcolor{silvercell} 88.02 & \cellcolor{silvercell} 0.290 & \cellcolor{silvercell} 24.18 & \cellcolor{silvercell} 0.214 \\
\specialrule{0.5pt}{0.25pt}{0.25pt}
{\small \fitnameshort}  & {\small MSD}   &
\cellcolor{goldcell}89.86 &
\cellcolor{goldcell}0.157 &
\cellcolor{goldcell}23.46 &
\cellcolor{goldcell}0.207 \\
\bottomrule
\end{tabular}
\caption{
Fitting and decomposition ablation:  
\fitnameshort, which interleaves analysis and optimization, outperforms a single-shot fitting approach (cf. Section~\ref{subsec:ablation}).  
Moreover, pairing \fitnameshort with MSD yields better results than using CoACD~\cite{coacd_wei_2022}.
}
\label{tab:decompose-ablation}
\end{table}
\paragraph{Timing.}
On the Toys4K test set, the full ten-round version of \fitnameshort~takes 652.6\,s per shape on average.
However, even a two-round variant offers a strong quality–time trade-off: it runs in 184.1\,s while achieving 86.54 IOU with only 15.54 primitives.  
This matches—and slightly exceeds—the reconstruction accuracy of MPS at $256^3$ resolution (86.30 IOU) while using \emph{over $5\times$ fewer} primitives and a comparable runtime (194.6\,s).  
PA (58.3\,s) and MPS (37.9\,s at $128^3$) are faster but produce lower-quality assemblies.  
We note that \fitnameshort\ is not yet optimized for speed; dedicated CUDA kernels for \primitivename\ may reduce runtime.

%% file: sections/05_applications.tex
\section{Applications}
\label{sec:applications}

Our representation enables several downstream uses that combine visual quality, editability, and analytic structure.  
We highlight four such applications and provide implementation details in the supplementary. 

\paragraph{Editable and Deployable Asset Generation.}
\label{sec:apps-assets}

\begin{figure}[t!]
	\centering
	\includegraphics[width=1.00\linewidth]{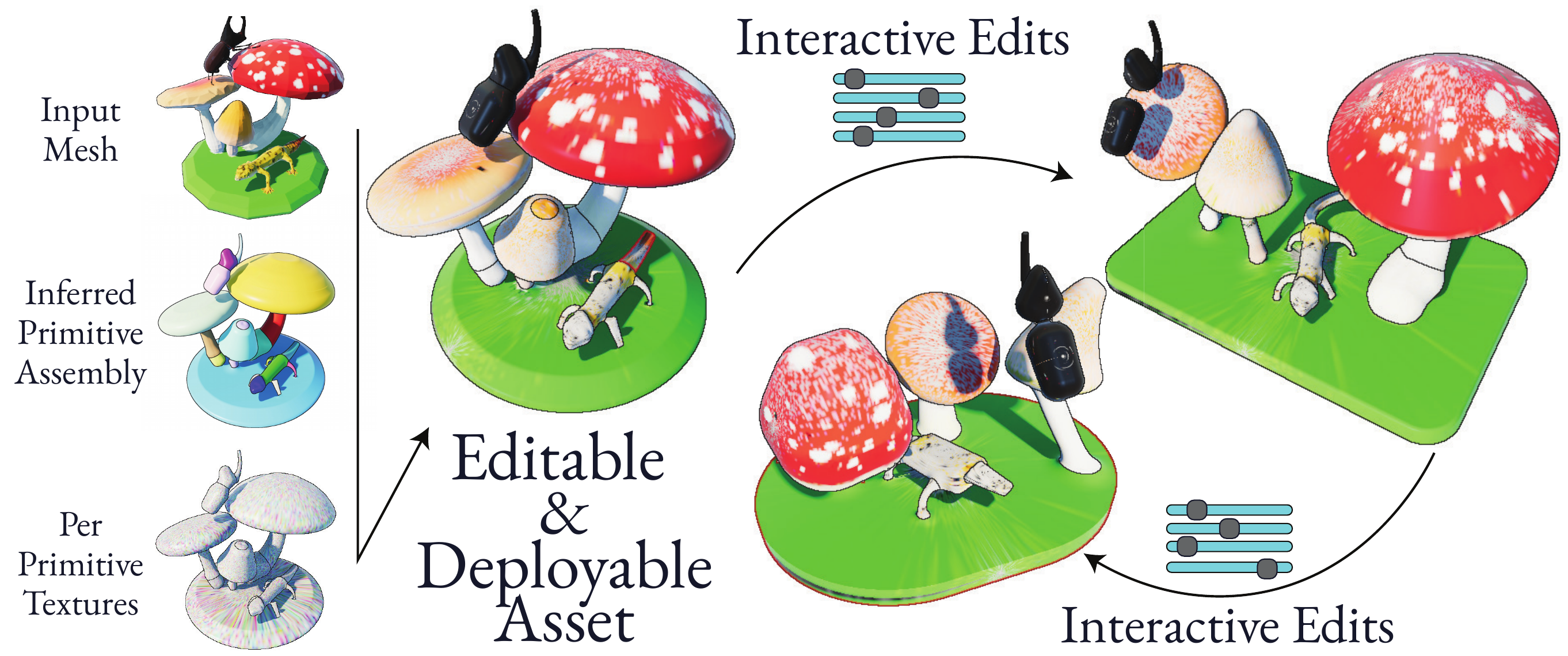}
	\caption{ 
    Assigning per-primitive spherical 2D textures \& optimizing it against a textured mesh, begets \textit{Edtiable} \& \textit{Deployable} assets.
 }
\label{fig:editing}
\end{figure}

Our primitives are simultaneously compact, editable, and capable of high-fidelity reconstruction, allowing them to serve directly as deployable 3D assets.  
To produce textured assemblies, we associate each primitive with a local 2D spherical texture map that we 
optimize 
against the target textured mesh.  The resulting textured assemblies can be directly deployed in real-time sphere traced scenes, while remaining  editable (see Fig.~\ref{fig:editing}).

\begin{table}[t]
\small
\centering
\setlength{\tabcolsep}{6pt}
\renewcommand{\arraystretch}{1.15}
\begin{tabular}{lcccc}
\toprule
\textbf{Method} &
{\footnotesize IOU} &
{\footnotesize CD}  &
{\footnotesize \#Prims}  &
{\footnotesize Overlap}  \\
\midrule
{\small CAPRI-Net}~\cite{caprinet_yu_2022}  &
\cellcolor{silvercell}84.46 &
\cellcolor{goldcell}0.100 &
23.0 & N/A \\
\specialrule{0.5pt}{0.25pt}{0.25pt}
{\small Ours (Solid)} &
82.64 &
0.192 &
\cellcolor{silvercell}14.18 &
0.342 \\
{\small Ours} &
\cellcolor{goldcell}88.30 &
\cellcolor{silvercell}0.127 &
\cellcolor{goldcell}13.16 &
\cellcolor{goldcell}0.186 \\
\bottomrule
\end{tabular}
\caption{
CSG inference on the ABC dataset~\cite{abc_Koch_2019_CVPR}: 
Our method achieves comparable reconstruction accuracy to CAPRI-Net~\cite{caprinet_yu_2022} while using significantly fewer primitives.
}
\label{tab:caprinet}
\end{table}

\paragraph{Inferring Canonical CSG Programs.}
\label{sec:apps-csg}

Although our framework is designed for smooth, soft-union assemblies, it can also infer discrete Constructive Solid Geometry (CSG) programs composed of canonical solids.  
We achieve this by constraining the parameters of \primitivename\ to be a barycentric interpolation of parameters to fetch canonical shapes—cuboid, cylinder, cone, and sphere—within the SuperFrustum formulation.  
Despite the lack of subtraction as a compositional operator, our primitive space natively contains ``subtracted'' shapes via the onion operator, which we include in the list of canonical shapes.
Fitting under these constraints yields solid CSG programs that remain compact and interpretable.  
In Table~\ref{tab:caprinet}, we compare this constrained version of our method, named ``solid''  to CAPRI-Net~\cite{caprinet_yu_2022} on a randomly sampled subset of the ABC dataset~\cite{abc_Koch_2019_CVPR} containing 100 samples.  
Our approach achieves nearly the same CSG reconstruction accuracy as CAPRI-Net while using roughly two-third as many primitives.
As shown in Fig.~\ref{fig:editing}, our inferred programs produce cleaner, less entangled assemblies—owing to the analytic expressivity of our primitives and the structured refinement in \fitnameshort.  

\paragraph{Image to primitive.}
\label{img2prim}

\begin{figure}[t!]
	\centering
	\includegraphics[width=1.00\linewidth]{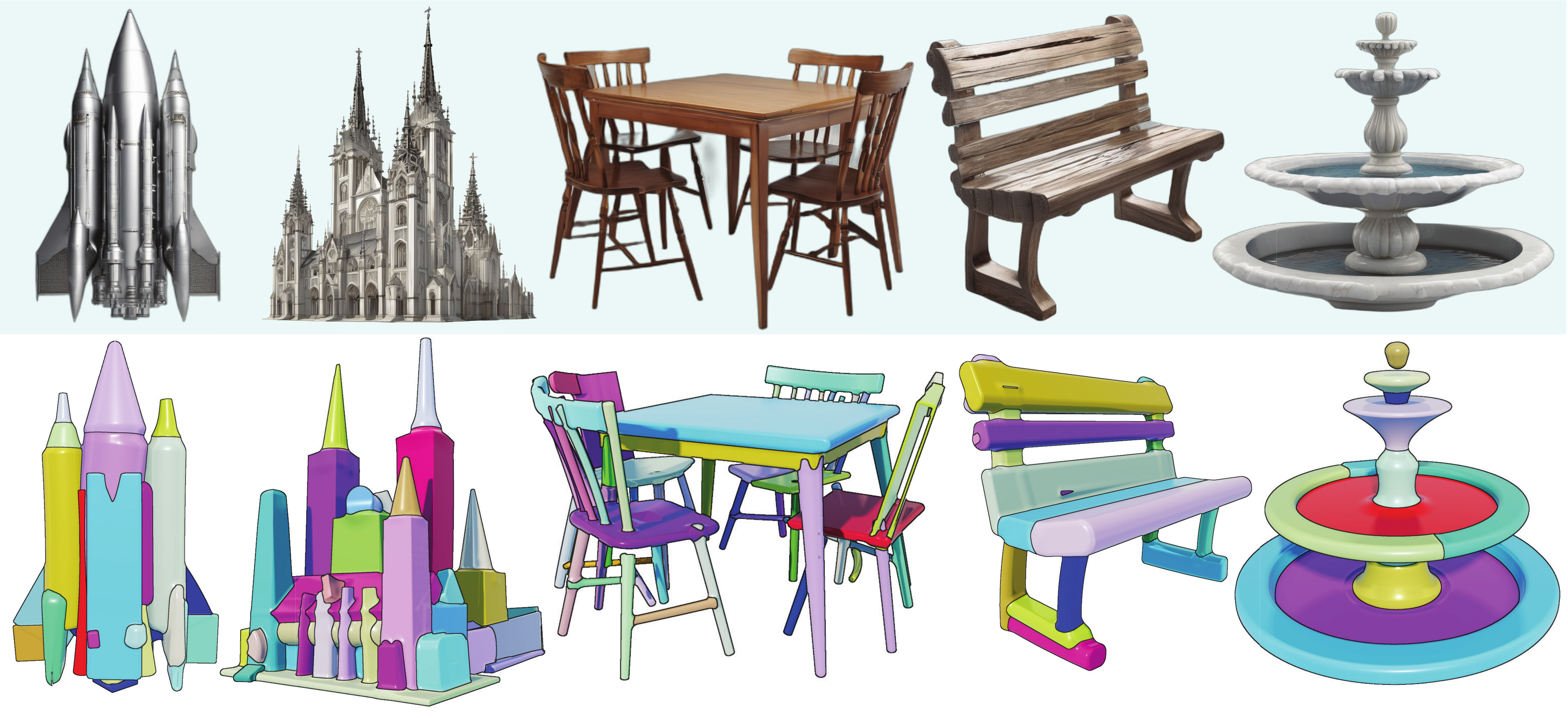}
	\caption{ 
    Our method can infer primitive assemblies from images by leveraging Text-2-3D models (Hunyuan3D-2.1~\cite{hunyuan3d_tencent_2025}).
 }
\label{fig:img2prim}
\end{figure}

Our method can seamlessly be combined with 3D generative models to achieve image to primitive assembly. In Figure~\ref{fig:img2prim}, we use Hunyuan3D-2.1~\cite{hunyuan3d_tencent_2025} with  \fitnameshort on samples from the 3DGen-Bench suite~\cite{3dgenbench_zhang_2025}.

\begin{figure}[t!]
	\centering
	\includegraphics[width=1.00\linewidth]{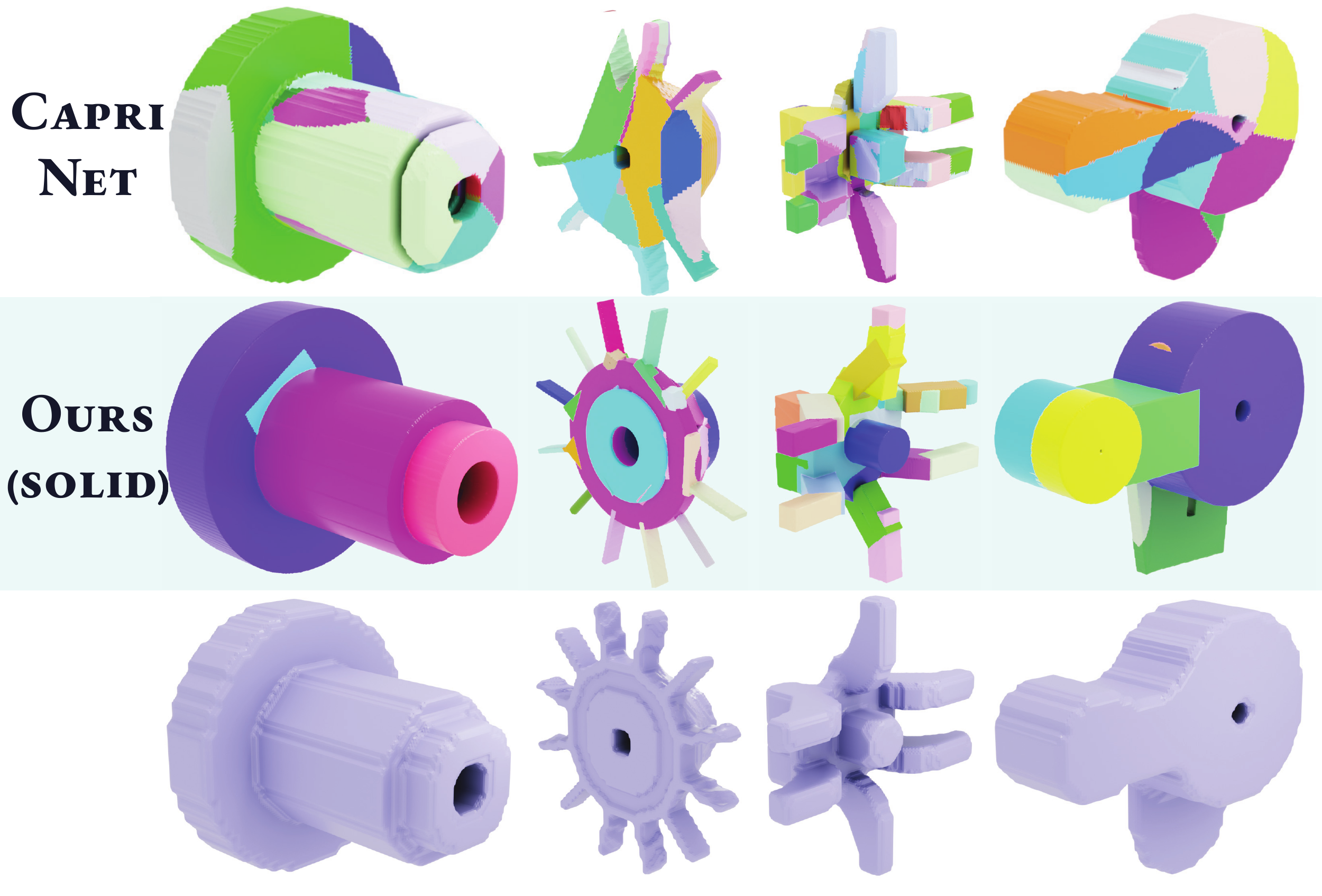}
	\caption{ 
    Our method can infer CSG programs using canonical solids (e.g., cylinders, cuboids; cf. Sec.~\ref{sec:apps-csg}), producing far more parsable and structured trees than CAPRI-Net~\cite{caprinet_yu_2022}.
 }
\label{fig:solid}
\end{figure}

\paragraph{Semantic Segmentation Enrichment.}
\label{sec:apps-seg}

Manually annotating parts can be quite expensive - as a result datasets often provide coarse grained annotations. 
ResFit can help to annotate finer parts. 
In Figure~\ref{fig:segmentation}, we intersect coarse semantic labels from the PartObjVerse dataset~\cite{partobjtiny_yang_2024sampart3d} with our primitive assemblies to enhance segmentation granularity. As a result, we subdivide large parts into functionally meaningful subcomponents without drifting outside their semantic boundaries.  This suggests a promising direction for integrating analytic decomposition as a prior for open-world part segmentation tasks.

%% file: sections/06_conclusions.tex
\section{Conclusion}

We introduced a framework for converting 3D shapes into compact and editable assemblies of analytic primitives.  
Our method combines two contributions: \primitivename, an expressive, compact, and optimizable analytic primitive; and \fitname, an iterative inference algorithm that couples shape analysis with primitive optimization to recover parsimonious yet accurate assemblies.  
Together, they shift the reconstruction–parsimony Pareto frontier, achieving state-of-the-art performance across benchmarks while producing high-fidelity, editable shape programs.
Despite its expressiveness, \fitnameshort\ is still restricted by its purely additive composition; shapes requiring subtractive operations remain challenging.  
Future work includes extending our decomposition strategies (e.g., tree-of-shapes) and developing richer applications in CSG modeling, interactive editing, and structured scene understanding.

\begin{figure}[t!]
	\centering
	\includegraphics[width=1.00\linewidth]{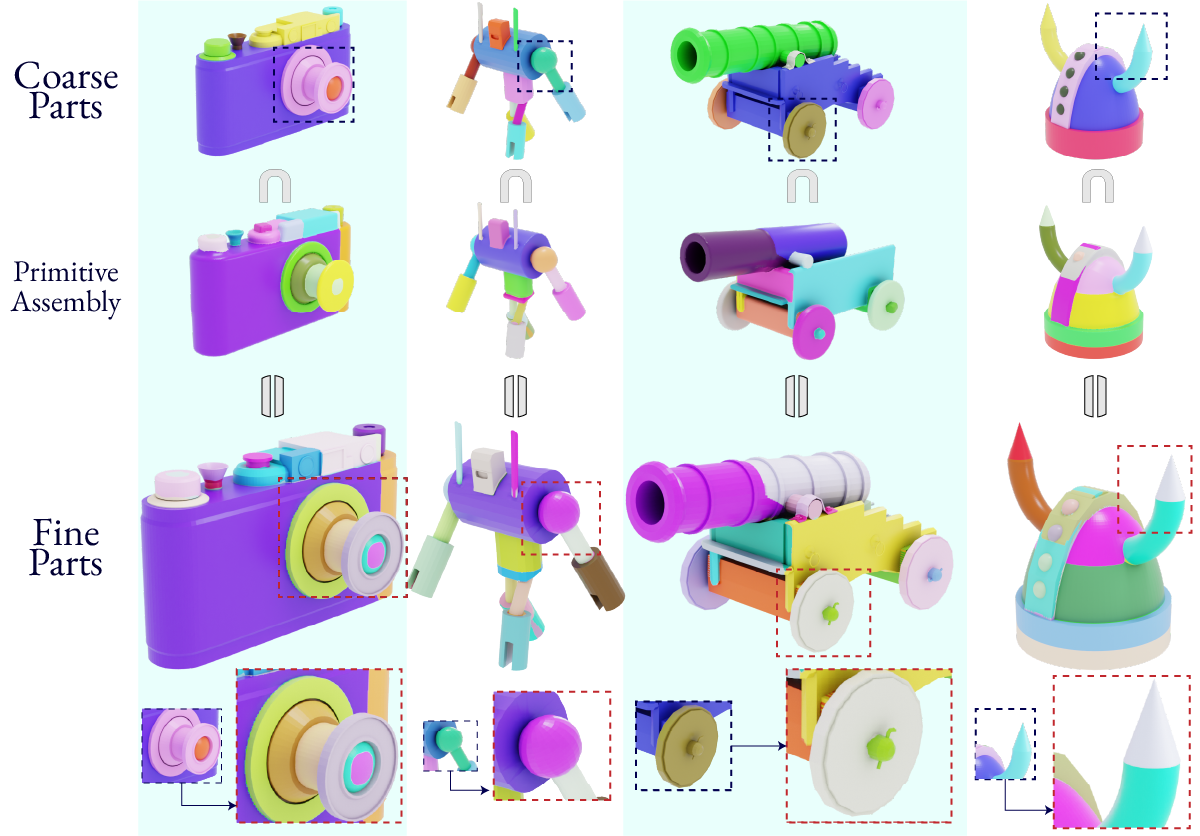}
	\caption{ 
ResFit enables finer, semantically consistent part segmentation.  
Row~1 shows the coarse semantic regions provided in PartObjVerse.  
Row~2 shows the primitive assemblies inferred by ResFit.  
Intersecting each coarse region with its corresponding assembly (Row~3) yields meaningful sub-parts—capturing functional structure while remaining strictly within the original semantic boundaries.  
 }
 \label{fig:segmentation}
\end{figure}